\documentclass[conference]{IEEEtran}
\IEEEoverridecommandlockouts

\usepackage[noadjust]{cite}
\usepackage{amsmath,amssymb,amsfonts}
\usepackage{algorithm}
\usepackage{graphicx}
\usepackage{adjustbox}

\usepackage{algpseudocode}

\usepackage{graphicx}
\usepackage{textcomp}
\usepackage{xcolor}
\usepackage{comment}
\usepackage{subfigure}
\usepackage{multirow}

\def\BibTeX{{\rm B\kern-.05em{\sc i\kern-.025em b}\kern-.08em
    T\kern-.1667em\lower.7ex\hbox{E}\kern-.125emX}}
\begin{document}

\title{Reuse Distance-based Copy-backs of Clean Cache Lines to Lower-level Caches}

\author{
\IEEEauthorblockN{Rui Wang, Chundong Wang$^*$\thanks{*Corresponding Author}, and Chongnan Ye}
\IEEEauthorblockA{\textit{School of Information Science and Technology}\\ 
\textit{ShanghaiTech University}\\
\{wangrui3, wangchd, yechn\}@shanghaitech.edu.cn
}}

\maketitle

\begin{abstract}
Cache plays a critical role in reducing the performance gap between CPU and main memory. 
A modern multi-core CPU generally employs a multi-level hierarchy of caches, through which
the most recently and frequently 
used data are maintained in each core's local private caches 
while all cores share the last-level cache (LLC).
For inclusive caches, clean cache lines replaced in higher-level caches
are not necessarily copied back to lower levels, as the inclusiveness implies their existences
in lower levels.
For exclusive and non-inclusive caches that
are widely utilized by Intel, AMD, and ARM today, 
either indiscriminately copying back all or none of replaced clean cache lines to lower levels
raises no violation to exclusiveness and non-inclusiveness definitions.

We have conducted a quantitative study and found that, 
copying back all or none of clean cache lines to lower-level cache 
of exclusive caches
entails suboptimal performance.
The reason is that only a part of cache lines would be reused
and others turn to be dead in a long run. 
This observation motivates us to selectively copy back some clean cache lines to LLC 
in an architecture of exclusive or non-inclusive caches. We revisit the concept of
{reuse distance} of cache lines.
In a nutshell, a clean cache line with a shorter reuse distance 
is copied back to lower-level cache as it is likely to be re-referenced in the near future,
while cache lines with much longer reuse distances would be discarded or 
sent to memory if they are dirty.
We have implemented and evaluated our proposal with 
non-volatile (STT-MRAM) LLC. Experimental results with gem5 and SPEC CPU 2017 benchmarks
show that on average our proposal 
yields up to 12.8\% higher throughput of IPC (instructions per cycle) than
 the least-recently-used (LRU) replacement policy with copying back all clean cache lines for STT-MRAM LLC.

\end{abstract}


\section{Introduction}

Cache is a critical component of CPU. It is employed to reduce the
performance gap between CPU and main memory. Modern CPUs generally
contain a multi-level hierarchy of caches.
Each CPU core has its private local caches that hold the most recently and frequently used data for itself 
while all cores share a larger last-level cache (LLC) that maintains data with less access recency and frequency.
There are three inclusion policies between multi-level caches.
For {\em inclusive} caches, cache lines in a higher-level cache are surely included in lower levels. In other words, a higher level is a subset of lower level.
In stark contrast, there is no intersection between multi-level {\em exclusive} caches.
 {\em Non-inclusive} caches do not enforce either aforementioned restriction and a cache line may or may not be found in any level.
 Non-inclusive and exclusive caches are more preferred by manufacturers such as Intel, AMD, and ARM,
 because inclusive caches introduce a wastage of space to redundantly keep duplicate 
 copies of the same cache line in multiple levels~\cite{yan2019attack}\cite{evenblij2019comparative}.

Owing to the limited capacity of cache,
from time to time,
some cache lines have to be selected as victims 
and replaced to make room for ones that are to be referenced. 
Although emerging non-volatile memory (NVM) technologies such as
STT-MRAM are supposed to provide much denser media for building larger
CPU caches in the near future~\cite{7851483}\cite{evenblij2019comparative}\cite{hameed2018performance}\cite{korgaonkar2018density}, cache replacement is still essential and imperative.An ineffectual replacement policy yet badly impairs performance. Assuming that
a cache line would be reused soon but erroneously replaced, the next reference to it causes a
cache miss and incurs a penalty of loading data from memory with considerable access time.
Multi-level caches help to reduce such penalties as lower levels can accept 
cache lines evicted from high levels, especially for dirty cache lines with modified data.
Mainstream multi-level caches mostly 
write back a dirty cache line into an immediately lower-level cache upon replacement 
rather than write it through multi-level caches and main memory.

Writing back dirty cache lines upon replacement is common in all inclusive, exclusive, and non-inclusive caches,
but how to handle replaced clean cache lines is not deterministic.
One straightforward way is to discard all such clean cache lines,
which does not violate the definitions of inclusiveness, exclusiveness, and non-inclusiveness.
Nevertheless, some of them may be reused soon, and discarding all is likely to incur cache misses with significant penalties.
On the other hand, exclusive caches in some CPUs
simply copy back\footnote{We use `copy back' for clean cache lines instead of `write back' because `write back' is conventionally used in literature to describe asynchronously updating dirty cache lines to 
lower levels of memory hierarchy. The latter is a contrast to the synchronous write-through strategy.} 
any clean cache line upon replacement to the immediately
lower-level cache~\cite{hameed2018performance}.

Factually, only a part of clean cache lines would be reused soon and other clean cache lines, as well as many dirty cache lines,
turn to be unused in a long run (known as dead cache lines~\cite{manivannan2016radar}\cite{gaur2011bypass}\cite{9407144}..
It is obvious that copying back all clean cache lines or none upon replacement is suboptimal, because 
the former underutilizes the lower-level cache while the latter may incur cache misses. We have done a quantitative study with exclusive caches for proof-of-concept, and found that, with the knowledge of future, selectively 
copying back cache lines that would be soon reused outperforms either copying all or none.

In the real world, predicting the future is mainly achieved by learning from the history.
We have considered an important historical record, i.e., the {\em reuse distance}, one that has been used to develop replacement policies,
to predict whether a cache line would be reused soon..
Cache lines with the longest reuse distances would be victims for replacement, but a clean cache line with a 
shorter reuse distance can be copied back into lower-level cache for a potential cache hit.
To summarize, we make following contributions in this paper.
\begin{itemize}
	\item We have conducted a quantitative study which indicates that, for exclusive or non-inclusive caches, copying back all clean cache lines or none yields inferior performance than partly copying back ones that would be reused soon.
	\item We measure the reuse distances of cache lines in the higher-level cache at runtime and predict whether
	to copy back a clean one into lower-level cache upon replacement.
	\item We confirm that dead cache lines exist in a larger lower-level cache. They are perfect victims to be replaced for placing copied-back clean cache lines.
\end{itemize}

We have done evaluation with gem5 simulator~\cite{gem5} by configuring two-level exclusive caches with non-volatile (STT-MRAM) LLC. Experimental results show that our proposal significantly improves performance compared to conventional LRU replacement with copying back all clean cache lines. For example, with
SPEC CPU\textsuperscript{\circledR} 2017 benchmarks~\cite{song2018experiments} run over a 1MB STT-MRAM L2 cache, our proposal yields up to
12.8\% higher throughput of IPC (instructions per cycle) on average.
 
 The remainder of the paper is as follows. Section~\ref{sec:motivation} outlines the background and the motivation of this paper, showing why we aim at copy-back prediction. Section~\ref{sec:ipml} presents our design CBP and its hardware implementation. Section~\ref{sec:eva} talks about the evaluation of our work on STT-MRAM L2 caches.  
 Section~\ref{sec:conclusion} concludes this paper.

\section{Background and Motivation}\label{sec:motivation}

\subsection{Overview of Multi-level Caches}
Cache provides data buffering 
between CPUs and memory. It contains a multi-level hierarchy in modern CPUs.
L1 cache is an on-core cache composed of instruction cache (ICache) and data cache (DCache). 
Lower-level caches hold more data with larger capacity. The LLC, with the largest capacity, is usually shared by multi-cores
to main data that are less frequently accessed than ones in each core's L1 cache.

Multi-level caches can be either inclusive, exclusive, or non-inclusive. The inclusive cache duplicates the cache line of the inner level, which helps coherence flows, but it reduces usable cache capacity. 
For exclusive caches, it is a benefit for capacity because no duplicate cache line exists between levels~\cite{yan2019attack}.
Non-inclusiveness is much more flexible compared to the former two and allows a cache line to stay at any level.
As a result, exclusive and non-inclusive caches are striding into mainstream CPUs and gradually
substitute inclusive caches as the de facto configuration today~\cite{lake}\cite{ryzen}.

No matter what inclusion policies a particular cache hierarchy uses, achieving effective cache utilization is an objective of primary importance. 
Prior studies~\cite{khan2010sampling}\cite{liu2008cache} found that
an efficient organization and management of data in the cache may cause
many clean cache lines, as well as many dirty lines, to become
 unused for a long time. They are known as dead cache lines and underutilize 
 the valuable resource provided by the cache, thereby leading to performance
 degradation.
 
Because of the limited capacity of caches, replacements are frequently triggered
to make room for cache lines that are to be accessed but missed. 
In a set-associative cache, 
the least-recently-used (LRU) cache line,
such as the aforementioned dead one, is usually the victim to be replaced from the cache set. 
Cache replacements happening at a higher level entail interactions with its immediate lower level.
Upon replacement, clean cache lines differentiate from dirty ones regarding the fact of
no change ever performed. For inclusive caches, both clean and dirty cache lines in a higher level
are surely existing in lower ones, so there is no need to push them back. 
As to exclusive and non-inclusive caches, clean cache lines to be replaced from high-level caches can be either discarded at all, or fully copied back into the lower level, as neither 
violates the definitions of exclusiveness and non-inclusiveness~\cite{yan2019attack}\cite{ARMmanual}.

\subsection{A Motivational Study on Copy-backs}\label{sec:bg}

We have conducted a study on the impact  
of copying back all or none of clean cache lines replaced
at high levels of a multi-level exclusive or non-inclusive cache hierarchy.
Hereafter we would use two-level exclusive caches for illustration unless otherwise stated. 
Similar observations are obtainable with more levels or non-inclusive caches.

\textit{Observation 1: Copying back all replaced clean cache lines into lower-level cache yields higher performance than discarding all of them}. 
We have first done a quantitative experiment with gem5 simulator in order to
illustrate the effect of various copy-back policies. 
We chose four representative benchmarks from the prevalent SPEC CPU\textsuperscript{\circledR} 2017 test suite.
The configurations of L1 DCache and L1 ICache were both 8-way set-associative with 32KB capacity. 
We set a 16-way set-associative L2 cache in 1MB STT-MRAM as the LLC. 
We applied the classic LRU algorithm for cache replacement.
By default, replaced dirty cache lines from L1 cache
 are filled into L2 cache.
As to replaced cache cache lines from L1 cache, 
we considered one of four following copy-back policies at runtime.
\begin{enumerate}
	\item Copying back clean cache line lines from both L1 DCache and ICache to L2 cache.
	\item Copying back none of clean cache lines from both L1 DCache and L1 ICache.
	\item Copying back clean lines of L1 ICache only to L2 cache.
	\item Copying back clean lines of L1 DCache only to L2 cache.
\end{enumerate}

\begin{figure}[t]
	\centerline{\includegraphics[width=0.5\textwidth]{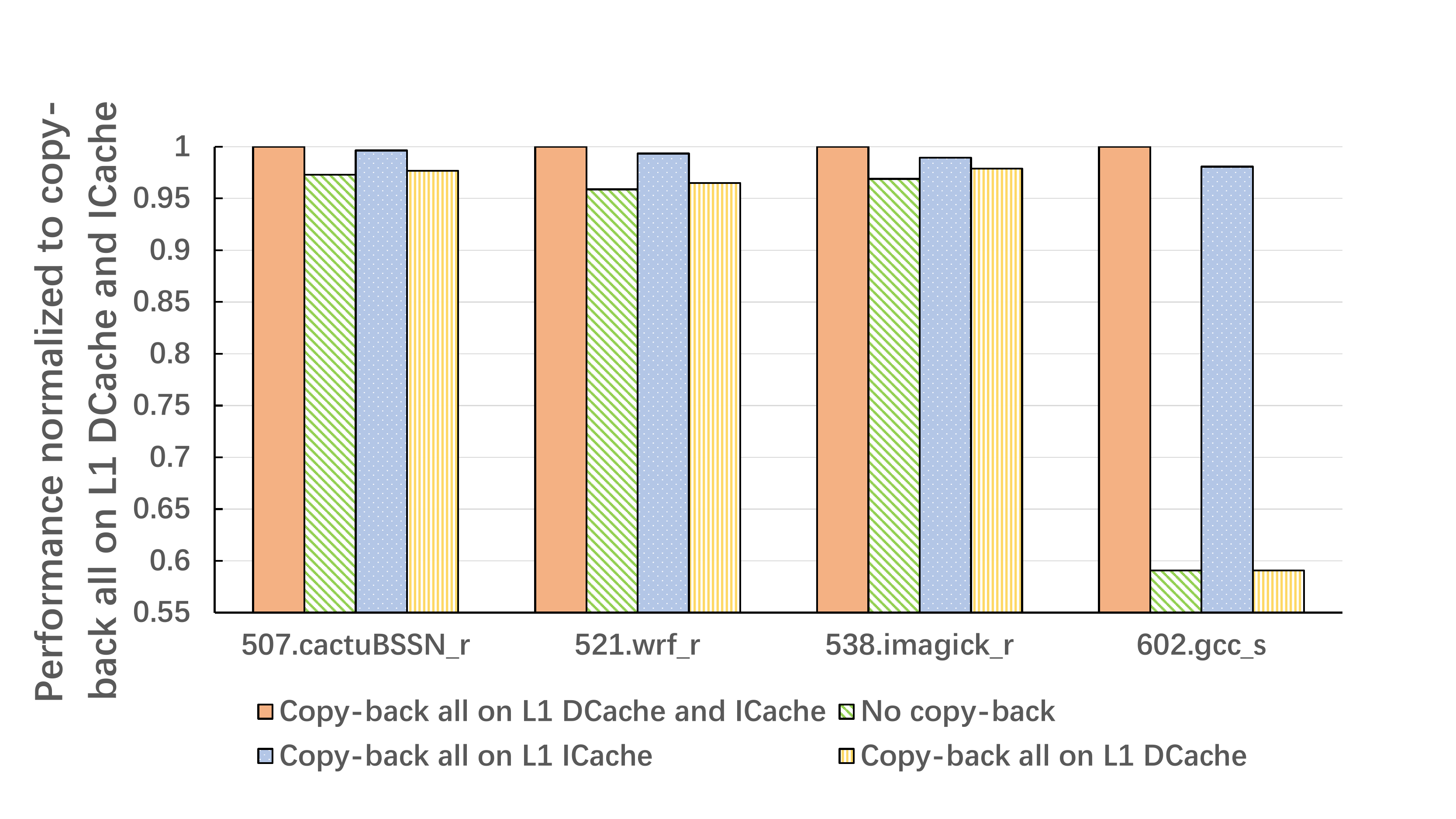}}
	\caption{IPC of various copy-back policies}
	\label{IPC}
\end{figure}

 Fig.~\ref{IPC} indicates
the throughputs (instructions per second, IPC) of four copy-back policies, normalized against a policy that copies back
all the clean cache lines to L2 cache. Fig.~\ref{IPC} clearly shows that copying back all clean cache lines achieves the highest
performance compared to the other policies. Furthermore, copying back none of clean cache lines attains the lowest performance among four.

\begin{figure}[t]
	\centerline{\includegraphics[width=0.5\textwidth]{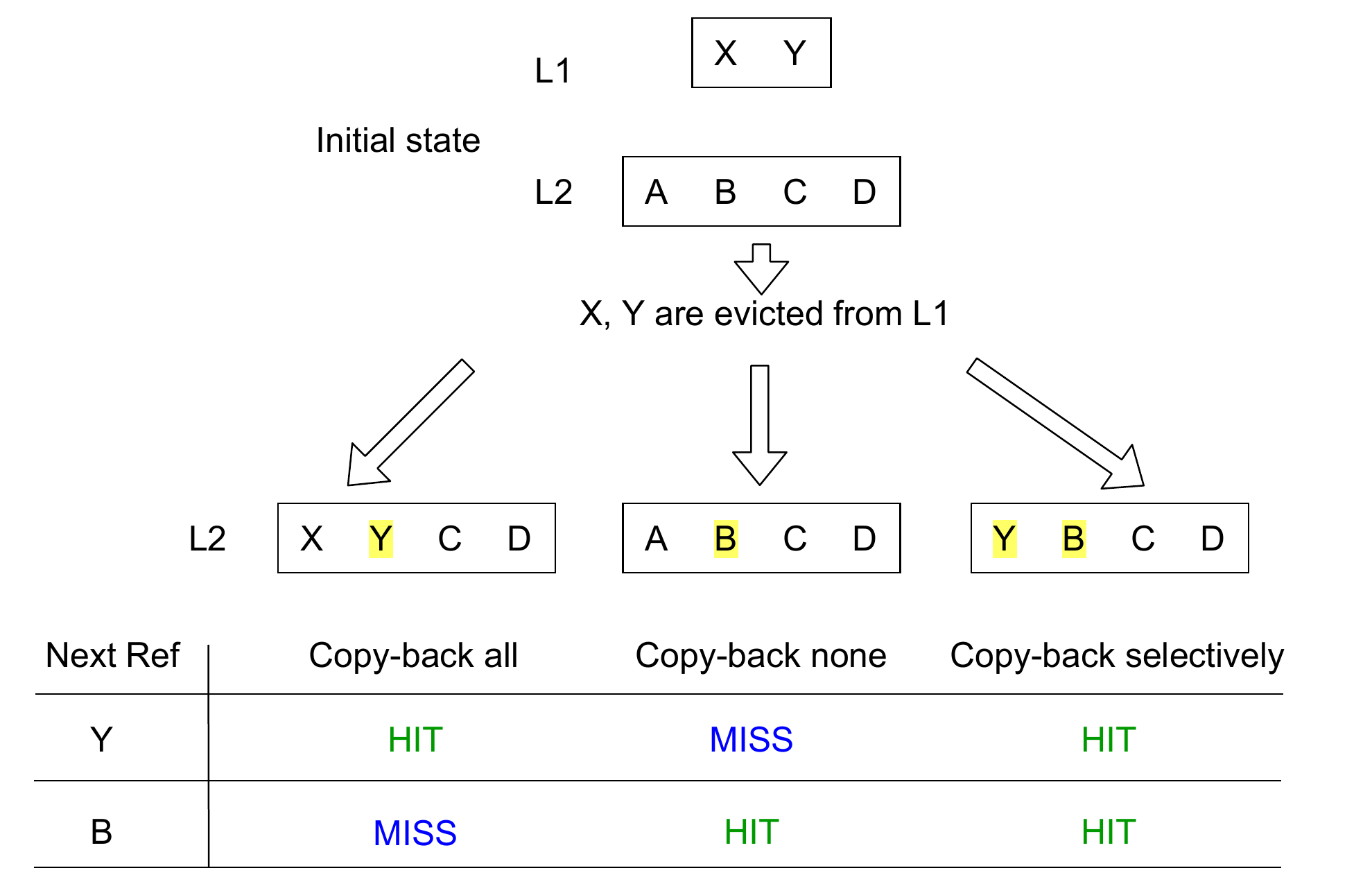}}
	\vspace{-1ex}
	\caption{The impact of copy-back policies on performance}
	\label{fig2}
\end{figure}

\begin{figure}[t]
	\centerline{\includegraphics[width=0.5\textwidth]{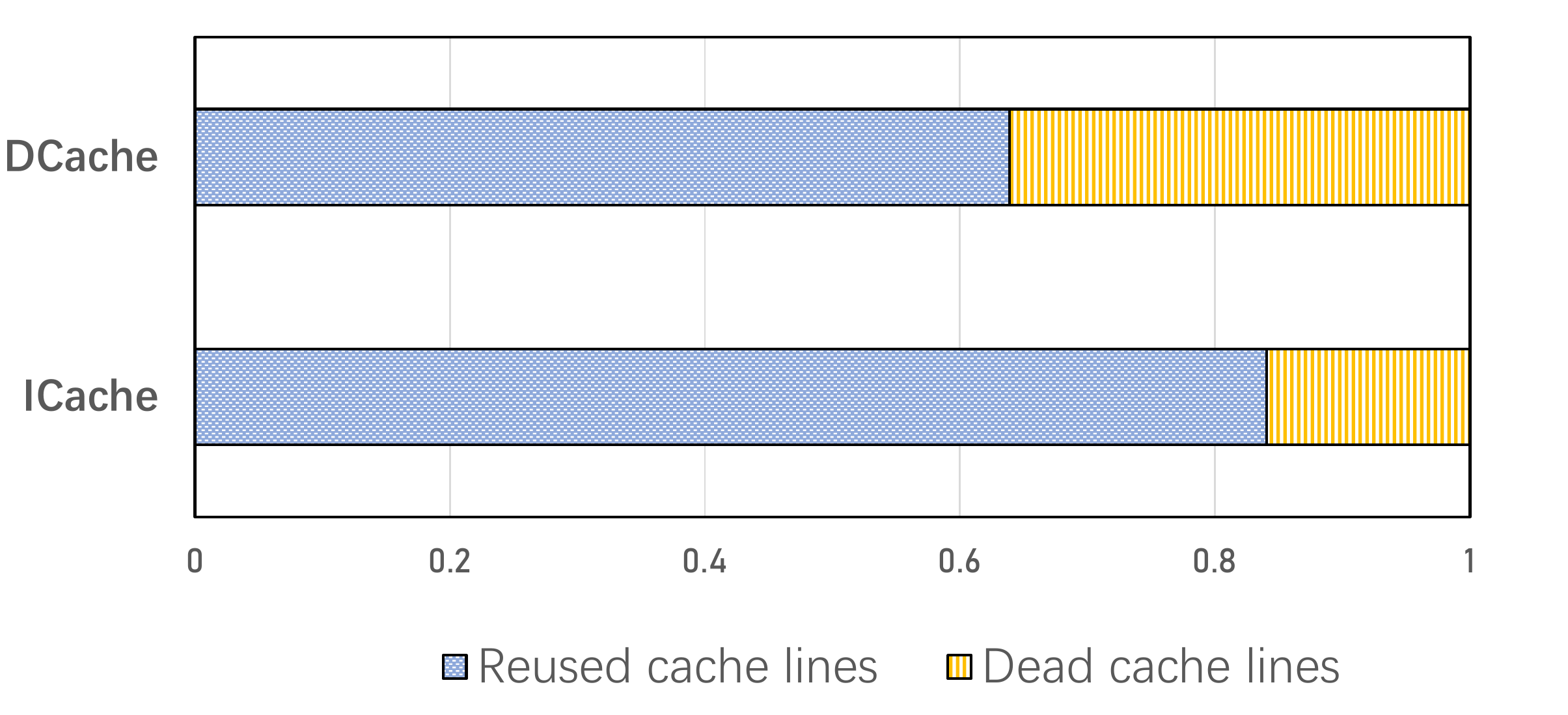}}
	\vspace{-1ex}
	\caption{The breakdown of cache lines with 602.gcc\_s}
	\label{reused_dist}
\end{figure}

\textit{Observation 2: copying back all clean cache lines yields suboptimal performance because 
some of them would not be reused in the near future}. 
We conducted another experiment with the two-level exclusive caches. 
We recorded the access history of L1 ICache and DCache and calculated 
the {\em reuse distance} of each cache line. We define the reuse distance
of a cache line as the number of misses over all cache lines in the cache 
between two consecutive
accesses to the specified one. For example,
the reuse distance of a cache line is 
five if globally five misses have happened before the next revisit of the cache line. 
We deem
a cache line to be dead if it is with a reuse distance greater than one thousand.
In Fig.~\ref{reused_dist}, both L1 ICache and DCache hold a number of dead cache lines, i.e., 16.0\% and 36.0\%, respectively.
Hence, copying back all clean cache lines, with dead ones included, yields
inferior performance, because dead cache lines are unlikely to be accessed again but the cost 
of copying them back and replacing existent ones in L2 cache for space is inevitable.
In fact, Fig.~\ref{IPC} shows that 
copying back all clean cache lines from L1 DCache entails much lower IPC than doing so with clean cache lines from L1 ICache.
The two bars of Fig.~\ref{reused_dist} explains the reason: 
L1 ICache contains fewer dead cache lines than DCache, since 
instructions are more frequently reused.

\textit{Observation 3: Selectively copying back some clean cache lines from higher level
to lower level improves performance.}
Let us take a simplified two-level exclusive cache hierarchy in Fig.~\ref{fig2} to do
 a qualitative analysis. 
The upper part of Fig.~\ref{fig2} shows the initial state of
two levels exclusively holding
six cache lines, i.e., \{X, Y\} in L1 and \{A, B, C, D\} in L2. We assume both X and Y are clean and they 
are successively evicted from L1 cache. Later CPU intends to access Y and B.
The lower part of Fig.~\ref{fig2} illustrates three possible situations with regard to copying back both X and Y, copying back neither X nor Y, and selectively copying back Y only.
It is evident that selectively copying back Y incurs fewer cache misses and in turn higher performance. The answer to the question on 
what clean cache lines to be {\em selected} for copying back
is the core of this paper.

\subsection{STT-MRAM as LLC}

In the last decade, multiple NVM technologies have been investigated. Some NVM technologies, such
as STT-MRAM, are promising to substitute SRAM for CPU caching, particulary as the LLC.
With higher density than SRAM,
STT-MRAM offers a large capacity cache design. Nonetheless, STT-MRAM suffers from longer write latency as well as higher write energy compared to SRAM. 
The long write latency of STT-MRAM leads to congestions on the request queues of LLC and can even stall CPU cores~\cite{korgaonkar2018density}. 
The avoidance of copying back clean cache lines that are dead to STT-MRAM LLC 
shall relieve the congestions for STT-MRAM, thereby improving the performance of STT-MRAM LLC.

\section{Design and Implementation}\label{sec:ipml}

\subsection{Design Decisions}

In this section, we would present our design of leveraging the reuse distance to select clean cache lines to be copied 
back from a high-level cache to a low-level one upon replacement.
The reuse distance has been widely used to develop cache replacement policies.
We find that the reuse distance is an appropriate indicator that
foresees future access to a cache line. 
In the current high-level cache, cache line with much longer reuse distances
are victims for replacement, but not all of them have a very long reuse distance
and would not be re-accessed in the near future.
Therefore, we embed our copy-back selection scheme in the replacement
process and set a threshold to filter out cache lines that are likely to
be re-referenced from ones that shall be discarded.

As mentioned, dead cache lines exist in all levels of a cache hierarchy.
Clean cache lines predicted to be reused soon can replace dead cache lines
in the lower level. This may cause write-backs to main memory, but
the gain in reducing future cache misses through selectively copying back cache lines offsets such penalties. 
There are research works utilizing the program counter (PC) to find out
dead cache lines mainly for LLC~\cite{khan2010sampling}\cite{kharbutli2008counter}\cite{liu2008cache}\cite{lai2001dead}.
Researchers also proposed PC-based replacement policies~\cite{wu2011ship}\cite{young2017ship++}. 
However, the involvement of PC in deciding whether to evict a cache line or
declare the death of it is unfavorable in practice, 
because a pipe forwarding the PC from CPU's front end to the memory hierarchy is 
hard to be designed, implemented, and verified. 
As a result, we have considered  
 integrating a non-PC-based replacement policy with the copy-back prediction for both
 high performance and verifiable hardware expenses.

\subsection{Design of CBP}

\paragraph*{Overview}
A state-of-the-art study~\cite{9407137} on cache behaviors using the reinforcement learning 
reveals that characteristics of a cache line, such as its data type (prefetched or loaded on cache miss), 
access frequency,
and reuse distance, are essentially 
contributive to performance. This inspires us of developing a copy-back predictor (CBP) based 
on these characteristics, especially on the reuse distance.
During the replacement process, our CBP closely collaborates with the replacement policy
to determine what cache lines to be replaced as victims and what victims shall be copied back.
We assign different cache lines with different {\em priorities}.
In particular,
 three types of cache lines have a higher priority to be replaced: 
 one with longer reuse distance, one
 with lower hit frequency and
 one containing prefetched data. 
For a clean cache line with a priority exceeding a predefined threshold,
 it is not copied back to the immediate lower-level cache.

\paragraph*{Private and Shared Reuse Distances} We maintain two types of reuse distances
for CBP. One is a private reuse distance for each valid cache line, initialized as
zero when the cache line is loaded into the high-level cache. 
We use {\em rd} to denote it.
{\em rd} is incremented by one every time a cache miss occurs to any arbitrary cache line,
and reset to be zero once a cache hit happens to the cache line {\em rd} stands for.
The other reuse distance is {\em RD}, an average reuse distance for all valid
cache lines in a cache set over time. 
In implementation, we use 4-bit saturating counter for both {\em rd} and {\em RD} (0 to 15).
Algorithm~\ref{alg:rule3} shows the calculation of them within a cache set.
On a cache miss, the {\em rd} of every valid cache line in the set is incremented by one (Lines 3 to 5).
Otherwise, the hit cache line's {\em rd} is reset and the {\em RD} of the cache set
is updated if eight hits have occurred to the set since the last update (Lines 10 to 14).
We note that the reason why we recalculate {\em RD} every eight cache hits is twofold.
On one hand, empirical studies done by us as well as by other researchers~\cite{9407144}
show that a shorter interval ($\le$8) incurs too frequent calculations while a
a longer internal ($\ge$8) hardly inflicts any difference to the value of $RD$.
On the other hand, doing an average over eight can be easily implemented as 
a shift-to-right operation.

\paragraph*{Other factors to be Considered} 
To improve the selection for replacement and copy-back, we 
also consider other factors besides the reuse distance. 
Our quantitative analysis finds that
the type of a cache line (prefetched or loaded on miss) and the hit frequency
of it impose an evident impact on cache replacement and copy-back.
We hence include them to calculate the eventual priority of each cache line.
As to the type of a cache line, due to the inaccuracy of the prefetcher,
prefetched cache lines have a high likelihood of ending up as the dead ones~\cite{lu2003performance}\cite{beyler2007performance}.
We give each cache line 1 bit to state whether it is prefetched or not.
As to the hit frequency of a cache line, the reuse distance in 
Algorithm~\ref{alg:rule3} does not
take into account such an important metric. It is widely acceptable that a cache line that
has been frequently accessed is probable to be accessed again. We
thus use a 2-bit saturating counter to record how many hits each cache line
has been accessed since the last replacement of a cache set.

\begin{algorithm}[t]
          \small
          \caption{Calculation of {\em rd} and {\em RD}}\label{alg:rule3}
          \begin{algorithmic}[1]
                  \Procedure{Calculate\_{\em rd}}{}\Comment{}
                  \If{it is a cache miss}  
                  \For{each valid cache line $cl$ in the set} 
                  \State  $cl$.{\em rd} $\leftarrow$  $cl$.{\em rd} + 1
                  \EndFor\label{endfor1}
                  \Else \Comment{$hit\_cl\ is\ the\ hit\ cache\ line$}
                  \State RDsum $\leftarrow$ RDsum + $hit\_cl$.{\em rd}
                  \State  $hit\_cl$.{\em rd} $\leftarrow$ 0 
                  \State RDcounter $\leftarrow$ RDcounter + 1
                  \If{RDcounter = 8} 
                  \State {\em RD} $\leftarrow$ RDsum / 8
                  \State RDsum $\leftarrow$ 0
                  \State RDcounter $\leftarrow$ 0
                  \EndIf\label{endif2}
                  \EndIf\label{endif3}
                  \EndProcedure
          \end{algorithmic}
\end{algorithm}

\paragraph*{The Computation of Priority} 
With the factors of reuse distance, being prefetched or not, and hit frequency for each cache line,
we can build a new replacement strategy with copy-backs.
Algorithm \ref{alg:RLR} describes how to compute the priority of each cache line upon a replacement request
among a cache set (Line 2 to Line 15). 
At the very beginning, each one has a priority of zero (Line 3).
A prefetched cache line, or a cache line has lower hit frequency, would have its priority
incremented by one, respectively. Next, the comparison between private {\em rd} amd the average {\em RD}
may give credit to the cache line (Line 10 to Line 14). More important, a greater {\em rd}
would entitle more credits (Line 11 and Line 13) as the cache line has not been accessed for a longer time.
We note that, due to the importance of reuse distance, the weight assigned to it is much higher than
the other two factors.

Then, a cache line with the highest priority is the victim to be evicted (Line 16). If the victim is dirty, or
it is clean but with a priority below a threshold,
it is copied back to the lower-level cache (Line 17 to Line 19).
We set the threshold for clean cache lines as nine, because the priority
of a cache line that is prefetched (1) or not frequently used (1), and not recently used (8)
be a sum of 9. Our extensive experiments also practically confirm the efficacy of using
nine as the threshold.
In the end, the victim is evicted from the current high-level cache.
We note that a strict order must be enforced between copying back the victim and evicting it.

\begin{algorithm}[t]
        \small
        \caption{The replacement with CBP in a cache set}\label{alg:RLR}
        \begin{algorithmic}[1]
                \Procedure{CBP}{}
                \For{each valid cache line $cl$ in the set}
                \State $cl$.priority $\leftarrow$ 0
                \If{IsPrefecthed($cl$) = True}
                \State $cl$.priority $\leftarrow$ $cl$.priority + 1
                \EndIf\label{endif4}
                \If{HitCounter($cl$) $\leq$ 1}
                \State $cl$.priority $\leftarrow$ $cl$.priority + 1
                \EndIf\label{endif5}
                \If{2 $\times$ {\em RD} $\leq$ {\em $cl$.rd} $\leq$ 3 $\times$  {\em RD}}
                \State $cl$.priority $\leftarrow$ $cl$.priority + 4
                \ElsIf{3 $\times$ {\em RD} $<$ {\em $cl$.rd}}
                \State $cl$.priority $\leftarrow$ $cl$.priority + 8
                \EndIf\label{endif7}
                \EndFor\label{endfor2}
                \State $victim$$ \leftarrow$ findHighestPriorityCL()
                \If{$victim$.priority $<$ 9 or IsDirty($victim$) = True}
                \State copyback($victim$)
                \EndIf\label{endif8}
                \State evict($victim$)
                \EndProcedure
        \end{algorithmic}
\end{algorithm}

\subsection{Hardware Cost}
For our proposal, we have added 7 bits ($rd$) for each cache line and a dozen of bits ($RD$, counter, etc.) 
per cache set. The common size of a cache line is 64B in modern commercial CPUs.
By calculation, the additional storage costs about 1.3\% of the total area. 
More important, the PC is not taken into consideration in our design, so the area cost as well as the design
and verification expenses are insignificant.

\section{Evaluation}\label{sec:eva}

\subsection{Methodology}

We implemented and evaluated the proposed replacement policy with CBP in the gem5 simulator on the
aforementioned two-level exclusive cache hierarchy.
Table~\ref{table:configure} shows the detailed configuration of the CPU we simulated.
We applied our proposed replacement policy with CBP onto the L1 DCache.
In particular, we experimented with STT-MRAM as the shared L2 cache (LLC).
We selected nine workloads from the SPEC CPU\textsuperscript{\circledR} 2017 benchmark suite, because
these benchmarks have high MPKI and are sensitive to varying cache organizations. 

The main metric to measure performance is the instructions per cycle (IPC). 
IPC reveals the efficiency and fluency of program execution. 
A higher IPC means a higher throughput.
We compared our proposal (denoted as \texttt{CBP}) to the classic LRU replacement policy with copying back all clean 
cache lines (denoted as \texttt{LRU}). For a clear illustration, 
all results of \texttt{CBP} are normalized against ones obtained with \texttt{LRU} unless otherwise stated.

\begin{table}[t]
\centering
\caption{The Configuration of Simulated CPU}
\label{table:configure}
\begin{adjustbox}{max width=\textwidth}
\begin{tabular}{|l|l|}
\hline
 Simulator & gem5 \\ \hline
 Benchmark & SPEC CPU 2017 \\ \hline
 CPU core & in order, 64-bit \\ \hline
 L1 ICache & 32KB, 8-way, LRU \\ \hline
 L1 DCache & 32KB, 8-way, CBP \\ \hline
\multirow{2}*{L2 Cache} 
               & 16-way STT-MRAM, 1 MB, LRU\\ 
               & read latency = 10 cycles, write latency = 40 cycles  \\ \hline

 Prefetcher & Stride prefetcher \\ \hline
 DRAM  &  8 GB DDR3-1600 \\ \hline
\end{tabular}
\end{adjustbox}
\end{table}

\begin{table*}[t]
	\centering
	\small
\caption{A Comparison on the Copy-backs between {\tt CBP} and {\tt LRU}}\label{table:copybacks}
	\begin{adjustbox}{width=1\textwidth}
		\begin{tabular}{|c|c|c|c|c|c|c|c|c|c|} 
			\hline
			\multicolumn{1}{|l|}{\begin{tabular}[c]{@{}l@{}}\# of copy-backs\\  for L1 DCache\end{tabular}} & 507.cactuBSSN\_r                  & 521.wrf\_r                       & 538.imagick\_r                 & 548.exchange2\_r             & 557.xz\_r                        & 600.perlbench\_s                 & 602.gcc\_s                       & 623.xalancbmk\_s                 & 631.deepsjeng\_s                  \\ \hline
			LRU                                                                                      & {\color[HTML]{333333} 16,023,511} & {\color[HTML]{333333} 6,032,447} & {\color[HTML]{333333} 328,764} & {\color[HTML]{333333} 5,833} & {\color[HTML]{333333} 2,050,938} & {\color[HTML]{333333} 1,960,125} & {\color[HTML]{333333} 2,359,821} & {\color[HTML]{333333} 8,843,999} & {\color[HTML]{333333} 71,275,938} \\ \hline
			CBP                                                                                      & 14,646,325                        & 5,261,716                        & 329,265                        & 5,832                        & 2,051,612                        & 1,945,706                        & 2,327,206                        & 8,840,750                        & 71,276,105                        \\ \hline
		\end{tabular}
	\end{adjustbox}
\end{table*}

\begin{figure}[htbp]
    \centering
    \includegraphics[width=0.5\textwidth]{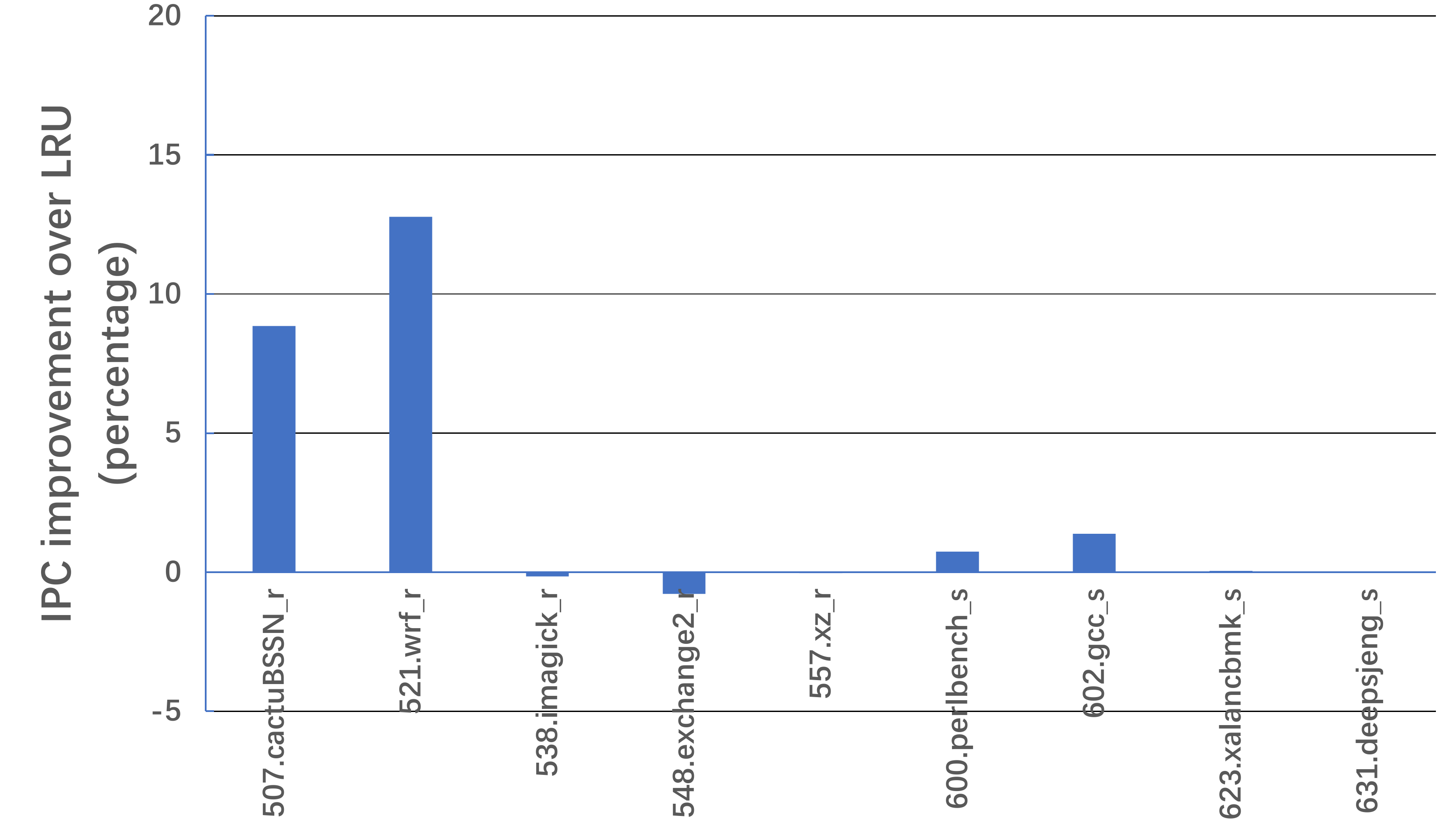}
    \caption{A comparison on IPC between {\tt LRU} and {\tt CBP}}
    \label{eval:IPC-STTRAM}
    \vspace{-2ex}
\end{figure}

\begin{figure}[t]
    \centering
    \includegraphics[width=0.5\textwidth]{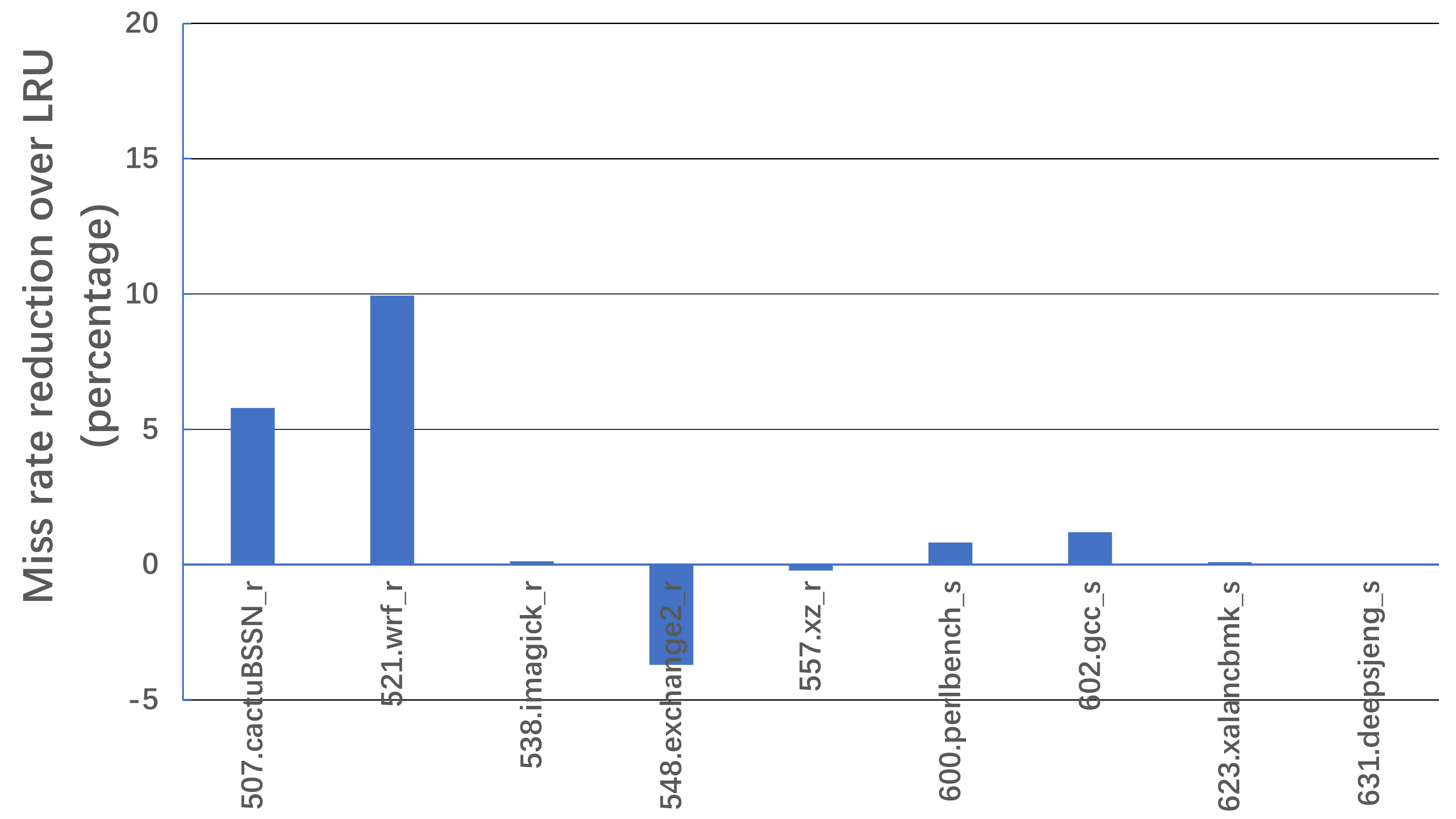}
    \caption{The reduction of miss rate by {\tt CBP} over {\tt LRU}}
    \label{eval:MR-STTRAM}
       \vspace{-1ex} 
\end{figure}

\subsection{Performance Evaluation}

We ran both {\tt CBP} and {\tt LRU} on STT-MRAM L2 cache to investigate their effects.
Fig.~\ref{eval:IPC-STTRAM} shows the IPC improvement of {\tt CBP} against {\tt LRU}.
It is evident that {\tt CBP} significantly outperforms {\tt LRU} with generally much higher
IPC. For example, with 521.wrf\_r, {\tt CBP} yields as much as 12.8\% more IPC than
{\tt LRU}. On average, the improvement of IPC achieved by {\tt CBP} is 2.5\%
compared to {\tt LRU} over all nine benchmarks. 
In addition, Fig.~\ref{eval:MR-STTRAM} shows the miss rate reduction of L1 DCache with both
{\tt CBP} and {\tt LRU}. We can obtain a similar observation between Fig.~\ref{eval:MR-STTRAM} and
Fig.~\ref{eval:IPC-STTRAM} and the miss rate is dramatically reduced by {\tt CBP}.
These results state that the idea of {\tt CBP}, i.e., leveraging
the reuse distance and other factors to filter out clean cache lines that
are going to be re-accessed and put them into L2 cache, is effectual.

We have recorded the numbers of copy-backs of both {\tt CBP} and {\tt LRU} after running
each benchmark. Table~\ref{table:copybacks} presents them.  
We note that {\tt LRU} copies back all while {\tt CBP}
selectively copies back some cache lines in accordance with Algorithm~\ref{alg:RLR}.
A joint investigation of Table~\ref{table:copybacks} and Fig.~\ref{eval:IPC-STTRAM} tells that fewer 
copy-backs lead to higher throughput. On average, {\tt CBP} copies back 2.9\%
 fewer cache lines than {\tt LRU}. For 521.wrf\_r, the reduction between {\tt CBP} and {\tt LRU}
 is 14.6\%. Selectively copying back some clean cache lines, rather than all of them like {\tt LRU},
 avoids keeping dead cache lines in lower-level cache, thereby improving the cache utilization.
 
In addition, from Fig.~\ref{eval:IPC-STTRAM}, we can find that, on the benchmark 548.exchange2\_r, {\tt CBP} is inferior compared
to {\tt LRU}. The numbers of copy-backs of {\tt CBP} and {\tt LRU} 
for this benchmark in Table~\ref{table:copybacks} explain the reason.
Compared to other benhmarks that would cause milions of or even tens of millions of
copy-backs, both {\tt CBP} and {\tt LRU} only did with about 5,800 cache lines, which is less in multiple orders of magnitude.
Such insignificant copy-backs hinder {\tt CBP} from outperforming {\tt LRU}.

\section{Conclusion}\label{sec:conclusion}

Exclusive and non-inclusive caches are becoming the mainstream configurations for a multi-level cache hierarchy in commericial CPUs. On the other hand,
NVM technologies such as STT-MRAM are promising candidates to be used as LLC. To make the most of multi-level caches,
we have studied 
the problem of efficiently handling clean cache lines replaced from higher-level caches.
By leveraging the reuse distance and other factors, we have proposed a new replacement policy with a selective copy-back
scheme for clean cache lines. Experimental results confirm that our proposal is able to
substantially outpeform the state-of-the-art with up to 12.8\% higher throughput.

{
\bibliographystyle{IEEEtran}	
\bibliography{reference}

\begin{thebibliography}{10}
\providecommand{\url}[1]{#1}
\csname url@samestyle\endcsname
\providecommand{\newblock}{\relax}
\providecommand{\bibinfo}[2]{#2}
\providecommand{\BIBentrySTDinterwordspacing}{\spaceskip=0pt\relax}
\providecommand{\BIBentryALTinterwordstretchfactor}{4}
\providecommand{\BIBentryALTinterwordspacing}{\spaceskip=\fontdimen2\font plus
\BIBentryALTinterwordstretchfactor\fontdimen3\font minus
  \fontdimen4\font\relax}
\providecommand{\BIBforeignlanguage}[2]{{%
\expandafter\ifx\csname l@#1\endcsname\relax
\typeout{** WARNING: IEEEtran.bst: No hyphenation pattern has been}%
\typeout{** loaded for the language `#1'. Using the pattern for}%
\typeout{** the default language instead.}%
\else
\language=\csname l@#1\endcsname
\fi
#2}}
\providecommand{\BIBdecl}{\relax}
\BIBdecl

\bibitem{yan2019attack}
M.~Yan, R.~Sprabery, B.~Gopireddy, C.~Fletcher, R.~Campbell, and J.~Torrellas,
  ``Attack directories, not caches: Side channel attacks in a non-inclusive
  world,'' in \emph{2019 IEEE Symposium on Security and Privacy (SP)}.\hskip
  1em plus 0.5em minus 0.4em\relax IEEE, 2019, pp. 888--904.

\bibitem{evenblij2019comparative}
T.~Evenblij, M.~Perumkunnil, F.~Catthoor, S.~Sakhare, P.~Debacker, G.~Kar,
  A.~Furnemont, N.~Bueno, J.~I. G{\'o}mez-P{\'e}rez, and C.~Tenllado, ``A
  comparative analysis on the impact of bank contention in {STT-MRAM and SRAM}
  based {LLCs},'' in \emph{2019 IEEE 37th International Conference on Computer
  Design (ICCD)}.\hskip 1em plus 0.5em minus 0.4em\relax IEEE, 2019, pp.
  255--263.

\bibitem{7851483}
Z.~Sun, X.~Bi, H.~Li, W.-F. Wong, Z.-L. Ong, X.~Zhu, and W.~Wu, ``Multi
  retention level {STT-RAM} cache designs with a dynamic refresh scheme,'' in
  \emph{2011 44th Annual IEEE/ACM International Symposium on Microarchitecture
  (MICRO)}, 2011, pp. 329--338.

\bibitem{hameed2018performance}
F.~Hameed, A.~A. Khan, and J.~Castrillon, ``Performance and energy-efficient
  design of {STT-RAM} last-level cache,'' \emph{IEEE Transactions on Very Large
  Scale Integration (VLSI) Systems}, vol.~26, no.~6, pp. 1059--1072, 2018.

\bibitem{korgaonkar2018density}
K.~Korgaonkar, I.~Bhati, H.~Liu, J.~Gaur, S.~Manipatruni, S.~Subramoney,
  T.~Karnik, S.~Swanson, I.~Young, and H.~Wang, ``Density tradeoffs of
  non-volatile memory as a replacement for {SRAM} based last level cache,'' in
  \emph{2018 ACM/IEEE 45th Annual International Symposium on Computer
  Architecture (ISCA)}.\hskip 1em plus 0.5em minus 0.4em\relax IEEE, 2018, pp.
  315--327.

\bibitem{manivannan2016radar}
M.~Manivannan, V.~Papaefstathiou, M.~Pericas, and P.~Stenstrom, ``Radar:
  Runtime-assisted dead region management for last-level caches,'' in
  \emph{2016 IEEE International Symposium on High Performance Computer
  Architecture (HPCA)}.\hskip 1em plus 0.5em minus 0.4em\relax IEEE, 2016, pp.
  644--656.

\bibitem{gaur2011bypass}
J.~Gaur, M.~Chaudhuri, and S.~Subramoney, ``Bypass and insertion algorithms for
  exclusive last-level caches,'' in \emph{Proceedings of the 38th annual
  international symposium on Computer architecture}, 2011, pp. 81--92.

\bibitem{9407144}
C.~Mazumdar, P.~Mitra, and A.~Basu, ``Dead page and dead block predictors:
  Cleaning tlbs and caches together,'' in \emph{2021 IEEE International
  Symposium on High-Performance Computer Architecture (HPCA)}, 2021, pp.
  507--519.

\bibitem{gem5}
``gem5 simulator,'' \url{https://www.gem5.org/}.

\bibitem{song2018experiments}
S.~Song, Q.~Wu, S.~Flolid, J.~Dean, R.~Panda, J.~Deng, and L.~K. John,
  ``Experiments with spec cpu 2017: Similarity, balance, phase behavior and
  simpoints,'' \emph{tech. rep., TR-180515-01}, 2018.

\bibitem{lake}
Intel, ``Intel core i5-10600k processor,'' 2021.

\bibitem{ryzen}
AMD, ``Ryzen 9 5900x desktop processors,'' 2021.

\bibitem{khan2010sampling}
S.~M. Khan, Y.~Tian, and D.~A. Jimenez, ``Sampling dead block prediction for
  last-level caches,'' in \emph{2010 43rd Annual IEEE/ACM International
  Symposium on Microarchitecture}.\hskip 1em plus 0.5em minus 0.4em\relax IEEE,
  2010, pp. 175--186.

\bibitem{liu2008cache}
H.~Liu, M.~Ferdman, J.~Huh, and D.~Burger, ``Cache bursts: A new approach for
  eliminating dead blocks and increasing cache efficiency,'' in \emph{2008 41st
  IEEE/ACM International Symposium on Microarchitecture}.\hskip 1em plus 0.5em
  minus 0.4em\relax IEEE, 2008, pp. 222--233.

\bibitem{ARMmanual}
\emph{\BIBforeignlanguage{English}{Cortex-A9 Technical Reference Manual}}, ARM.

\bibitem{kharbutli2008counter}
M.~Kharbutli and Y.~Solihin, ``Counter-based cache replacement and bypassing
  algorithms,'' \emph{IEEE Transactions on Computers}, vol.~57, no.~4, pp.
  433--447, 2008.

\bibitem{lai2001dead}
A.-C. Lai, C.~Fide, and B.~Falsafi, ``Dead-block prediction \& dead-block
  correlating prefetchers,'' in \emph{Proceedings 28th Annual International
  Symposium on Computer Architecture}.\hskip 1em plus 0.5em minus 0.4em\relax
  IEEE, 2001, pp. 144--154.

\bibitem{wu2011ship}
C.-J. Wu, A.~Jaleel, W.~Hasenplaugh, M.~Martonosi, S.~C. Steely~Jr, and
  J.~Emer, ``Ship: Signature-based hit predictor for high performance
  caching,'' in \emph{Proceedings of the 44th Annual IEEE/ACM International
  Symposium on Microarchitecture}, 2011, pp. 430--441.

\bibitem{young2017ship++}
V.~Young, C.-C. Chou, A.~Jaleel, and M.~Qureshi, ``Ship++: Enhancing
  signature-based hit predictor for improved cache performance,'' in \emph{The
  2nd Cache Replacement Championship (CRC-2 Workshop in ISCA 2017)}, 2017.

\bibitem{9407137}
S.~Sethumurugan, J.~Yin, and J.~Sartori, ``Designing a cost-effective cache
  replacement policy using machine learning,'' in \emph{2021 IEEE International
  Symposium on High-Performance Computer Architecture (HPCA)}, 2021, pp.
  291--303.

\bibitem{lu2003performance}
J.~Lu, H.~Chen, R.~Fu, W.-C. Hsu, B.~Othmer, P.-C. Yew, and D.-Y. Chen, ``The
  performance of runtime data cache prefetching in a dynamic optimization
  system,'' in \emph{Proceedings. 36th Annual IEEE/ACM International Symposium
  on Microarchitecture, 2003. MICRO-36.}\hskip 1em plus 0.5em minus 0.4em\relax
  IEEE, 2003, pp. 180--190.

\bibitem{beyler2007performance}
J.~C. Beyler and P.~Clauss, ``Performance driven data cache prefetching in a
  dynamic software optimization system,'' in \emph{Proceedings of the 21st
  annual international conference on Supercomputing}, 2007, pp. 202--209.

\end{thebibliography}
}

\end{document}